\documentclass[twocolumn,showpacs,preprintnumbers,amsmath,amssymb,APSl,prd,nofootinbib,superscriptaddress]{revtex4-1}

\usepackage{dcolumn}
\usepackage{bm}
\usepackage{ifpdf}
\usepackage{hyperref}
\usepackage{dcolumn}
\usepackage{bm}
\usepackage[spanish,english]{babel}
\usepackage{amsfonts}
\usepackage{amssymb}
\usepackage{graphicx}
\usepackage{color}
\usepackage[latin1]{inputenc}

\newcommand{\LL}{\mathcal{L}}

\newcommand{\be}{\begin{equation}}
\newcommand{\en}{\end{equation}}
\newcommand{\bea}{\begin{eqnarray}}
\newcommand{\ena}{\end{eqnarray}}

\begin{document}

\title{Melvin Universe in Born-Infeld gravity}

\author{Cosimo Bambi}\email{bambi@fudan.edu.cn}
\affiliation{Center for Field Theory and Particle Physics and Department of
Physics, Fudan University, 220 Handan Road, 200433 Shanghai, China}
\author{Gonzalo J. Olmo} \email{gonzalo.olmo@csic.es}
\affiliation{Departamento de F\'{i}sica Te\'{o}rica and IFIC, Centro Mixto Universidad de
Valencia - CSIC. Universidad de Valencia, Burjassot-46100, Valencia, Spain}
\affiliation{Departamento de F\'isica, Universidade Federal da
Para\'\i ba, 58051-900 Jo\~ao Pessoa, Para\'\i ba, Brazil}
\author{D. Rubiera-Garcia} \email{drubiera@fudan.edu.cn}
\affiliation{Center for Field Theory and Particle Physics and Department of
Physics, Fudan University, 220 Handan Road, 200433 Shanghai, China}


\date{\today}

\begin{abstract}
We consider a magnetic flux pointing in the $z$ direction of an axially symmetric space-time (Melvin Universe) in a Born-Infeld-type extension of General Relativity (GR) formulated in the Palatini approach. Large magnetic fields could have been produced in the early Universe, and given rise to interesting phenomenology regarding wormholes and black hole remnants. We find a formal analytic solution to this problem that recovers the GR result in the appropriate limits. Our results set the basis for further extensions that could allow the embedding of pairs of black hole remnants in geometries with intense magnetic fields.
\end{abstract}

\maketitle

\section{Introduction}

The Melvin Universe is a regular, non-black hole electrovacuum solution of the Einstein-Maxwell equations describing a bundle of magnetic flux lines in static equilibrium, held together by their own gravitational interaction \cite{Melvin}. It represents an explicit realization of Wheeler's geons \cite{Wheeler}, namely, self-consistent sourceless gravitational-electromagnetic entities  \cite{Misner}. This solution has been generalized to rotating and time-dependent configurations~\cite{Garfinkle94}, dilatonic~\cite{dilaton} and axion fields~\cite{axion}, higher dimensions~\cite{Higher}, and nonlinear electrodynamics~\cite{NED}. Its applications include supergravity~\cite{SUGRA} and D-Branes~\cite{Branes}.

The magnetic background of the Melvin solution can be employed to investigate the creation of black hole pairs, since the negative energy of the magnetic field compensates the energy of the pair and thus energy conservation is fulfilled~\cite{Pair}. If the magnetic field is strong enough, a Wheeler wormhole solution, namely, a pair of extreme Reissner-Nordstr\"om black holes identified at their throats and with opposite charges, can be formed \cite{wormholes}. This construction faces us with the problematic issue of topology change processes, in which the quantum foam scenarios are based. In the case of the Melvin space-time, such a process is governed by tunelling effects through instantons, which can be modeled in terms of Ernst's metric~\cite{Ernst}.

Large-scale coherent magnetic fields with strength of order $10^{-6}$~G are observed in galaxies and galaxy clusters~\cite{mf1}. According to the primordial hypothesis, these magnetic fields were created in the early Universe and they were later amplified by a dynamo mechanism (see e.g. \cite{Kunze} for a review). Production of microscopic black holes/wormholes in the early Universe by large fluctuations was first discussed by Hawking \cite{Hawking}, with the result that large numbers of such objects with Planck-order mass and a few units of charge could have been produced. Though the interest on such an idea faded away after the discovery of Hawking radiation, which would imply that such objects should have evaporated by the current epoch, recent research in extensions of GR using different approaches has put again regular black holes, wormholes, and black hole remnants under an intense discussion (see \cite{Chen} for a review). Indeed, a number of scenarios have been proposed~\cite{mf2} and some of them have the necessary ingredients for the production of microwormholes. Astrophysical observations put some constraints on the presence of wormholes in our Universe, but their existence cannot be ruled out~\cite{newref1}. For instance, some traversable wormholes are viable candidates to explain the supermassive objects at the centers of galaxies, while other kinds of wormholes can be excluded~\cite{Bambi}.

The main goal of this paper is to work out a Melvin-type space-time in an extension of GR which has gained interest in the last few years, namely, Born-Infeld gravity. Recently, it was shown~\cite{or13,or12} that this theory contains static, spherically symmetric wormholes supported by the electromagnetic field with geonic properties. A very important question is to discuss plausible mechanisms for the generation of such wormholes. An important step in this sense was done in \cite{lor14}, where it was shown that dynamical generation of wormholes through charged fluxes of radiation is possible. In this work we progress in a different, though related, direction hoping to clarify if such wormholes could be generated (or embedded) in highly magnetized scenarios. The first step requires to determine if Melvin-type solutions in analytical form can be found in this framework. We find that the answer is partially positive, as two analytical approximations to the exact solution can be found explicitly in the two regimes of interest. The fact of having some analytical control on the solutions suggests that an answer to the problem posed here might be accessible, though further research is still necessary.

The paper is organized as follows: in Sec. \ref{sec:BI}, we briefly review Born-Infeld gravity in the Palatini approach. In Sec. \ref{s-melvin}, we work out the axially symmetric Melvin-type space-time in this scenario, and obtain analytical solutions in the two regions of interest, namely, close to the axis and far from it. Conclusions and future research are discussed in Sec. \ref{s-c}.

\section{Action and main equations} \label{sec:BI}

In the last few years an extension of GR, that has attracted much attention in astrophysical and cosmological scenarios~\cite{BIapplications}, has been proposed following the analogy with the Born-Infeld theory of non-linear electrodynamics \cite{BIm}. Initially introduced in the metric formalism \cite{Deser}, the interest in this theory arose once its Palatini version was considered \cite{Banados}, as it avoids higher-order derivatives and ghosts. The action of this Born-Infeld theory of gravity (BI for short) coupled to an electromagnetic field can be written as
\bea \label{eq:BI}
S^{BI}&=&\frac{1}{\kappa^2 \epsilon} \int d^4x \left[ \sqrt{-q} - \lambda \sqrt{-g } \right] \nonumber \\
&-& \frac{1}{16\pi} \int d^4x \sqrt{-g} F_{\mu\nu}F^{\mu\nu} \, .
\ena
where $\kappa^2$ is Newton's constant in suitable units (in GR, $\kappa^2=8\pi G_{\rm N}$), $g$ and $q$ are the determinant of the space-time metric $g_{\mu\nu}$ and of the {\it metric} $q_{\mu\nu} \equiv g_{\mu\nu} + \epsilon R_{\mu\nu}$, with $\epsilon$ a small constant with dimensions of length squared, and $R_{\mu\nu}(\Gamma) \equiv {R^\alpha}_{\mu \alpha \nu}$ is the (symmetric) Ricci tensor constructed with the affine connection $\Gamma \equiv \Gamma^{\lambda}_{\mu\nu}$, which is a priori independent of the metric $g_{\mu\nu}$ (metric-affine or Palatini approach). In the matter (Maxwell) sector, $F_{\mu\nu}=\partial_{\mu}A_{\mu} - \partial_{\nu} A_{\mu}$ is the field strength tensor of the vector potential $A_{\mu}$. The meaning of $\lambda$ in Eq.~(\ref{eq:BI}) follows from a series expansion in powers of $\epsilon$ of the gravitational sector, $S_G$, as
\begin{equation} \label{eq:fseries}
S_G \approx \int  \frac{d^4x \sqrt{-g}}{2\kappa^2} \left[R-2\Lambda -\frac{\epsilon}{2} \left(-\frac{R^2}{2} + R_{\mu\nu}R^{\mu\nu}\right)+\ldots\right] \nonumber \, .
\end{equation}
This theory recovers GR with a cosmological constant term $\Lambda=\frac{\lambda-1}{\epsilon}$ at zeroth order, while higher-order corrections in the curvature invariants are suppressed by powers of $\epsilon$.

In the Palatini formulation of modified gravity, the usual troubles with higher-order field equations and ghost-like instabilities are avoided for large families of models. This is in sharp contrast with the situation in most approaches to modified gravity, where the connection is taken to be \emph{a priori} given by the Christoffel symbols of the metric. While in the special case of the Einstein-Hilbert action both approaches give the same Einstein equations, this is not so for most extensions of GR \cite{Borunda}. We point out that some experimental results regarding systems with defects in solid state physics seem to support both the Palatini approach and the Born-Infeld gravity action \cite{Volovich}, with potential consequences for our understanding of the microscopic description of space-time and gravitational phenomena \cite{lor-crystal}.

To implement the Palatini formalism, we perform independent variations of the action~(\ref{eq:BI}) with respect to the metric and the connection, which yields (setting torsion to zero for simplicity \cite{Torsion})
\begin{eqnarray}
\frac{\sqrt{\vert q \vert}}{\sqrt{\vert g \vert }} q^{\mu\nu}-\lambda g^{\mu\nu}&=&-\kappa^2 \epsilon T^{\mu\nu} \, , \label{eq:metric} \\
\nabla_{\alpha}^{\Gamma} \left(\sqrt{q} q^{\mu\nu} \right)&=&0 \, . \label{eq:connection}
\end{eqnarray}
From Eq.~(\ref{eq:connection}), it follows that the independent connection is given by the Christoffel symbols of the metric $q_{\mu\nu}$. Using (\ref{eq:metric}), one finds that $q_{\mu\nu}$ is related to the space-time metric $g_{\mu\nu}$ (for which one has $\nabla_{\alpha}^{\Gamma} (\sqrt{-g} g^{\mu\nu}) \neq 0$) as
\begin{equation} \label{eq:h-g}
\hat{q}=\sqrt{\vert \hat{\Sigma} \vert } \hat{\Sigma}^{-1} \hat{g} \, , \quad
\hat{q}^{-1}= \frac{ \hat{g}^{-1} \hat{\Sigma} }{\sqrt{ \vert \hat{\Sigma} \vert }}
\end{equation}
with the definition
\begin{equation}
\hat{\Sigma}  = \lambda \hat{I} - \epsilon \kappa^2 \hat{T} \, ,
\end{equation}
where $\hat{T} \equiv T^{\mu\alpha} g_{\alpha \nu}$ and a hat denotes a matrix. This clearly shows that the relation between $q_{\mu\nu}$ and $g_{\mu\nu}$ is algebraic and only governed by the matter fields.

From the definition $\hat{q}=\hat{g}+\epsilon \hat{R}$ and the relations~(\ref{eq:h-g}), a bit of algebra allows us to write the metric field equations~(\ref{eq:metric}) in terms of $q_{\mu\nu}$ as
\be \label{eq:fequations}
{R_\mu}^{\nu}(q)=\frac{\kappa^2}{\sqrt{\vert \hat{\Sigma} \vert}} \left( \LL_G {\delta_\mu}^{\nu} +  {T_\mu}^{\nu} \right) \, ,
\en
where $L_G= (\vert \hat{\Sigma} \vert^{1/2} - \lambda)/(\epsilon \kappa^2)$ is the gravity Lagrangian. This represents a set of second-order differential Einstein-like field equations with all the right-hand-side only depending on the matter sources. Since $q_{\mu\nu}$ is algebraically related to $g_{\mu\nu}$, the field equations for it are second-order as well. In vacuum, the field equations~(\ref{eq:h-metric}) boil down to
\be
R_{\mu\nu} (q)=\frac{(\lambda-1)}{\lambda\epsilon} q_{\mu\nu} \leftrightarrow R_{\mu\nu} (g)=\frac{(\lambda-1)}{\epsilon} g_{\mu\nu} \, . \nonumber
\en
This clearly shows that the dynamics of this theory in vacuum is that of GR with a cosmological constant term with value $\Lambda\equiv \frac{(\lambda-1)}{\epsilon}$. In addition, it implies that the theory is free of extra propagating degrees of freedom and ghost-like instabilities.

\section{Axial magnetic space-time} \label{s-melvin}

\subsection{Basic equations}

The electromagnetic (Maxwell) field in action (\ref{eq:BI}) satisfies the equations
\be \label{eq:em}
\nabla_{\mu}F^{\mu\nu}=0 \ \Leftrightarrow \ \partial_\mu\left(\sqrt{-g}F^{\mu\nu}\right)=0 \, .
\en
For a radial, static, and spherically symmetric field, the only non-vanishing component is $F^{tr} \equiv E(r)$. In this case, one finds that if $\epsilon=-2l_\epsilon^2$, with $l_\epsilon$ representing a length scale, then the point-like singularity of GR is generically replaced by a wormhole structure, whose properties have been studied in detail in a number of papers \cite{or12,or13}. We will thus adopt from now on the choice $\epsilon=-2l_\epsilon^2$.  Our interest here is to find the analog solution for the case of cylindrically symmetric and magnetically charged configurations in coordinates $(t,z,\rho,\phi)$. The physical interpretation is that of a space-time sourced by a beam of magnetic fields parallel to the $z$-axis. In order to solve the system of field equations~(\ref{eq:fequations}) for BI gravity~(\ref{eq:BI}) with an axially symmetric electromagnetic field described by Maxwell equations~(\ref{eq:em}), we introduce a line element for the metric $g_{\mu\nu}$ as follows
\be \label{eq:g-metric}
ds^2=f(\rho)(-dt^2+dz^2)+g(\rho)d\rho^2+h(\rho)\rho^2d\varphi^2 \, .
\en
With this line element, from the field equations~(\ref{eq:em}) we find that the only non-vanishing component of the field strength tensor is $F^{\rho \varphi}$, which satisfies
\be
\partial_\rho\left(\sqrt{-g}F^{\rho \varphi}\right)=0 \ \Rightarrow \ F^{\rho \varphi}=\frac{\beta}{\rho f \sqrt{gh}} \ ,
\en
where $\beta$ is an integration constant that determines the intensity of the magnetic field. With this result, one can verify that the field invariant is $F_{\mu\nu}F^{\mu\nu}=2F_{\rho\varphi}F^{\rho\varphi}=2 (F^{\rho\varphi})^2 g h \rho^2$. Accordingly, to get rid of the numerical factor, we define $X=-\frac{1}{2}F_{\mu\nu}F^{\mu\nu}=-\beta^2/f^2$. We note that $X$ depends neither on $g$ nor on $h$, which simplifies the analysis. With these results, we find that the energy-momentum tensor of the electromagnetic field
\be
{T_\mu}^{\nu}=-\frac{1}{4\pi} \left( {F_\mu}^{\alpha}{F_\alpha}^{\nu} - \frac{1}{4}{\delta_\mu}^{\nu}F_{\alpha\beta}F^{\alpha\beta} \right)
\en
takes the simple form
\be \label{eq:Tmn}
{T_\mu}^{\nu}=\frac{X}{8\pi} \text{diag}(1,1,-1,-1) \ .
\en
The matrix $\hat{\Sigma}$ of the theory, $\hat{\Sigma}=\hat{I}+2l_\epsilon^2\kappa^2 \hat{T}$, then becomes
\be \label{eq:Sigma}
{\Sigma_{\mu}}^{\nu}=\left(
\begin{array}{cc}
\sigma_- \hat{I} & \hat{0} \\
\hat{0} & \sigma_+ \hat{I}  \\
\end{array}
\right) \, ,
\en
where we have defined
\be
\sigma_{\pm}=1\pm \frac{f_c^2}{f^2} \, .
\en
To simplify the notation, here we have denoted $f_c^2\equiv\tilde{\kappa}^2 \beta^2 l_\epsilon^2$, with $\tilde{\kappa}^2\equiv \kappa^2/4\pi$, and $\hat{0}$ and $\hat{I}$ are the zero and identity $2 \times 2$ matrices, respectively. Note that $\tilde{\kappa}^2 \beta^2$ can be interpreted as an inverse squared length associated to the magnetic field, which we denote $l_\beta^2=1/\tilde{\kappa}^2 \beta^2$. This leads to $f_c=l_\epsilon/l_\beta$ and indicates that the ratio of the gravitational length scale $l_\epsilon$ to the magnetic length scale $l_\beta$ controls the departures of our solutions from those of GR.

Since the field equations~(\ref{eq:fequations}) will be solved in terms of the metric $q_{\mu\nu}$ associated to the affine structure, we define an axially symmetric line element for it as
\be \label{eq:h-metric}
d\tilde{s}^2= \tilde{f}(\rho)(-dt^2+dz^2) + \tilde{g}(\rho)d\rho^2 + \tilde{h}(\rho) \rho^2 d\phi^2 \, ,
\en
which is formally identical to that of the metric $g_{\mu\nu}$ [see Eq.~(\ref{eq:g-metric})]. From the relations~(\ref{eq:Sigma}), we immediately see that the functions defining the line elements of the metric $q_{\mu\nu}$~(\ref{eq:h-metric}) and $g_{\mu\nu}$~(\ref{eq:g-metric}) are related by
\be \label{eq:f-h-g}
\tilde{f}=\sigma_+ f \, ,  \,\,\, \tilde{h}=\sigma_- h \, ,  \,\,\, \tilde{g}=\sigma_{-}g \, .
\en
From the first of the above relations, it is easy to see that
\be\label{eq:f-ft}
f=\frac{\tilde{f} + \sqrt{\tilde{f}^2-4f_c^2}}{2} \ ,
\en
which indicates that $\tilde{f}$ is bounded by $\tilde{f}\ge 2f_c$. For this value of $\tilde{f}$, we get $f=f_c$, which corresponds to $\sigma_{-}=0$.

\subsection{Computation of the metric components and the line element}

With the elements above, we note that the BI Lagrangian can be written as
\be
\LL_G=\frac{\sqrt{ \det \hat{\Sigma}}-1}{-2{\kappa}^2 l_\epsilon^2}=\frac{\beta^2f_c^2}{8\pi f^4} \, .
\en
It is a simple calculation to show that the field equations~(\ref{eq:fequations}) for these axially symmetric solutions take the explicit form
\be\label{eq:Rmn(q)}
{R_\mu}^{\nu}(q)=-\frac{\tilde{\kappa}^2\beta^2}{2f^2} \left(
\begin{array}{cc}
\frac{1}{\sigma_+} \hat{I} & \hat{0} \\
\hat{0} & -\frac{1}{\sigma_-} \hat{I}  \\
\end{array}
\right) \, .
\en
Using the xAct package for Mathematica \cite{JMMG}, the components of ${R_\mu}^{\nu}(q)$ corresponding to the line element~(\ref{eq:h-metric}) can be easily obtained, and are given by
\bea
{R_0}^0&=& {R_1}^1 = -\frac{1}{4\tilde{f}^2} \left[-\frac{\tilde{f}_{\rho}^2}{\tilde{f}} +\tilde{f}_{\rho}\left(\frac{2}{\rho}+\frac{\tilde{h}_{\rho}}{\tilde{h}}\right)+2\tilde{f}_{\rho\rho} \right] \, \\
{R_2}^2&=&\frac{1}{4\tilde{f}^2\tilde{h}^2 \rho} \left[4\tilde{h}^2\rho \frac{\tilde{f}_{\rho}^2}{\tilde{f}}+\tilde{h}\left(\rho \tilde{f}_{\rho}\tilde{h}_{\rho}+2\tilde{h}\{\tilde{f}_{\rho}-2\rho\tilde{f}_{\rho\rho}\}\right) \right. \nonumber \\  &+& \left.\tilde{f}\left(\rho\tilde{h}^2_{\rho}-2\tilde{h}(2\tilde{h}_{\rho}+\rho \tilde{h}_{\rho\rho})\right)\right] \, , \\
{R_3}^3&=&\frac{1}{4\tilde{f}^2\tilde{h}^2 \rho} \Big[\tilde{f}\Big(\rho\tilde{h}^2_{\rho}-2\tilde{h}(2\tilde{h}_{\rho}+\rho \tilde{h}_{\rho\rho})\Big)   \nonumber \\
&-&\tilde{h}\Big(\rho \tilde{f}_{\rho}\tilde{h}_{\rho}+2\tilde{h}\tilde{f}_{\rho}\Big)\Big] \, ,
\ena
where we have imposed the gauge freedom $\tilde{g}=\tilde{f}$, and have used a subindex to denote derivative with respect to $\rho$. Given the block-diagonal form of the field equations~(\ref{eq:Rmn(q)}), it follows that ${R_2}^2-{R_3}^3=0$, which implies that
\be
0=\frac{1}{4\tilde{f}^2\tilde{h}^2 \rho} \left[4\tilde{h}^2\rho \frac{\tilde{f}_{\rho}^2}{\tilde{f}}+\tilde{h}\left(2\rho \tilde{f}_{\rho}\tilde{h}_{\rho}+4\tilde{h}\{\tilde{f}_{\rho}-\rho\tilde{f}_{\rho\rho}\}\right)\right] \, .
\en
With elementary algebraic manipulations, this equation becomes
\be\label{eq:R22-R33}
\left(\frac{\tilde{h}_\rho}{\tilde{h}}+\frac{2}{\rho}+\frac{2\tilde{f}_{\rho}}{\tilde{f}}\right)=\frac{2\tilde{f}_{\rho\rho}}{\tilde{f}_\rho} \ ,
\en
which can be readily integrated to obtain
\be \label{eq:h}
\tilde{h}\rho^2=\alpha \left(\frac{\tilde{f}_{\rho}}{\tilde{f}}\right)^2 \ ,
\en
where $\alpha$ is an integration constant (with dimensions), whose value we keep undefined for the moment.

Using now~(\ref{eq:R22-R33}) in the expression for ${R_0}^0$, we get the following equation for $\tilde{f}$:
\be \label{eq:tilde_f_0}
\tilde{f}_{\rho\rho}-\frac{3}{4}\frac{\tilde{f}^2_\rho}{\tilde{f}}=\frac{\tilde{\kappa}^2\beta^2}{2}\frac{\tilde{f}}{f} \ .
\en
We will next proceed to solve this equation to obtain $\tilde{f}(\rho)$. With the solution, the geometry $g_{\mu\nu}$ can be obtained using the relations~(\ref{eq:h-g}).

\subsection{Finding a solution}

Using the expression (\ref{eq:f-ft}) in (\ref{eq:tilde_f_0}), the equation to solve takes the form
\be \label{eq:tilde_f_1}
\tilde{f}_{\rho\rho}-\frac{3}{4}\frac{\tilde{f}^2_\rho}{\tilde{f}}=\frac{\tilde{\kappa}^2\beta^2\tilde{f}}{\tilde{f}+\sqrt{\tilde{f}^2-4f_c^2}} \ .
\en
This equation can be written in a more convenient (dimensionless) form by defining $\tilde{f}=2f_c \phi(x)$ and $\rho^2=\frac{2f_c}{\tilde{\kappa}^2\beta^2}x^2$, which leads to
\be \label{eq:tilde_f_2}
\phi_{xx}-\frac{3}{4}\frac{\phi^2_x}{\phi}=\frac{\phi}{\phi+\sqrt{\phi^2-1}} \ .
\en
We now define  a new function $\Omega=\phi_x^2$ such that $d\Omega/d\phi=2\phi_{xx}$, which turns (\ref{eq:tilde_f_2}) into
\be \label{eq:tilde_f_3}
\Omega_\phi-\frac{3}{2\phi}\Omega=\frac{2\phi}{\phi+\sqrt{\phi^2-1}} \ .
\en
This equation admits an exact solution for $\Omega(\phi)$ of the form
\begin{eqnarray}
\Omega&=&C \phi ^{\frac{3}{2}}+\frac{4 \phi ^2}{3} \left(\phi -\sqrt{\phi ^2-1}\right)\nonumber \\
&-& \frac{8 \phi}{3}\  _2F_1\left(\frac{1}{4},\frac{1}{2};\frac{5}{4};\frac{1}{\phi ^2}\right) \ ,
\end{eqnarray}
where $C$ is an integration constant and $_2F_1$ a hypergeometric function. Given that $\Omega=\phi_x^2$ must be a positive function by construction, an expansion about $\phi\approx 1$ indicates that $C$ cannot be arbitrary. In fact, we find that
\begin{eqnarray}
\lim_{\phi\to 1} \Omega&\approx& C+\frac{4}{3}-\frac{2 \sqrt{\pi } \Gamma \left(\frac{5}{4}\right)}{\Gamma \left(\frac{7}{4}\right)}\\
&+& (\phi -1) \left(\frac{3 C}{2}+4-\frac{4 \sqrt{\pi } \Gamma \left(\frac{5}{4}\right)}{\Gamma \left(\frac{3}{4}\right)}\right)+O[(\phi-1)^2] \nonumber \ ,
\end{eqnarray}
which requires that the constant factor when $\phi=1$ be positive or zero. Now, given that $\phi$ must be greater or equal to $1$ to guarantee the reality of the differential equation (\ref{eq:tilde_f_2}), on physical grounds we must demand that $\phi=1$ be a minimum. This forces us to set this constant to zero, which implies $C=-\frac{4}{3}+\frac{2 \sqrt{\pi } \Gamma \left(\frac{5}{4}\right)}{\Gamma \left(\frac{7}{4}\right)}\approx 2.16274$.  We note that in GR the constant $C$ is usually normalized as $C_{GR}=2$.

The expression for $\phi(x)$ can be obtained numerically by integrating $\phi_x=\Omega^{1/2}$ with the value of $C$ chosen above. It is useful, however, to find approximate analytical solutions in the two regimes of interest, namely when $\phi\to 1$ and when $\phi\gg 1$. The limit $\phi\gg 1$ is particularly interesting because the equation for $\phi(x)$ converges to the equation that one finds in GR. In fact, in this limit (\ref{eq:tilde_f_2}) turns into
\be \label{eq:tilde_f_GR}
\phi_{xx}-\frac{3}{2}\frac{\phi^2_x}{\phi}\approx 1 \ ,
\en
and proceeding as above, we get $\Omega_{GR}=C \phi^{3/2}-2\phi$ (where $C\phi^{3/2}$ represents the solution of the homogeneous equation). With this solution one can readily find $\phi(x)=4 (1 + (C x)^2/32)^2/C^2$, which is the right GR Melvin solution \cite{Melvin}. The choice $C_{GR}=2$ arises naturally and makes the metric Minkowskian as $x\to 0$, where $\phi(x)\to 1$. In our case, however, $C\neq 2$ is motivated by the behavior of the differential equation near $\phi\to 1$. In this other limit, we find that
\begin{equation}
\phi_x^2\approx 2(\phi-1) \ ,
\end{equation}
which leads to $\phi(x)\sim  1+x^2/2$. With this result, one can verify that the line element as $x\to 0$ takes the form
\begin{equation}
\frac{ds^2}{f_c}\approx (1+x)\left[-dt^2+dz^2+\rho_0^2\frac{dx^2}{x}\right]+\frac{\alpha}{2f_c\rho_0^2}x d\varphi^2 \ ,
\end{equation}
where $\rho_0^2\equiv 2l_\epsilon l_\beta$ and $\alpha$ is the (unspecified) integration constant that arose in (\ref{eq:h}). Under a rescaling of the $x$ coordinate of the form $dx^2/x= dy^2$, this line element becomes
\begin{equation}
\frac{ds^2}{f_c}\approx\left[-dt^2+dz^2+\frac{\rho_0^2}{4}{dy^2}\right]+\frac{\alpha}{8f_c\rho_0^2}y^2 d\varphi^2 \ ,
\end{equation}
where second-order corrections in $y^2$ have been neglected. This last expression shows that the Minkowski space-time can be recovered near the axis by just defining  $r=\rho_0 y/2$ and $\alpha\equiv 2f_c \rho_0^4$. An additional global rescaling of units could be used to absorb the constant factor $f_c$.

\begin{figure}[t]
\includegraphics[width=0.5\textwidth]{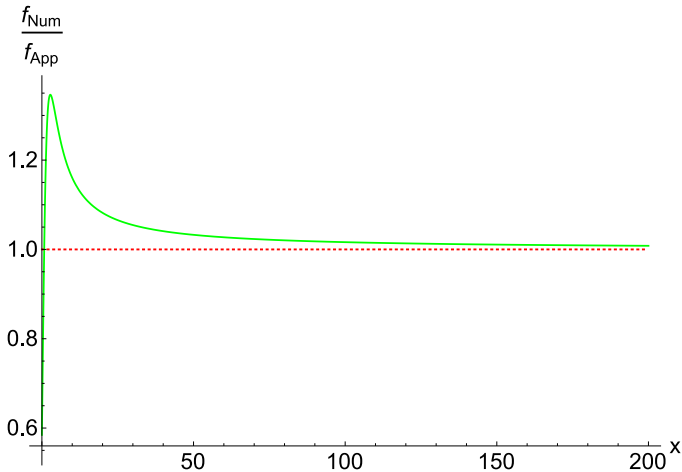}
\caption{Ratio of the function $f(x)$ computed numerically by its analytic approximation used in (\ref{eq:ds2far}). \label{fig:Ratio}}
\end{figure}

In the $\phi\gg 1$ limit, we find that $f\approx 2f_c\phi$. Taking the solution for $\phi$ obtained above in this limit, $\phi(x)=4 (1 + (C x)^2/32)^2/C^2$, the line element can be approximated as
\begin{eqnarray}\label{eq:ds2far}
\frac{ds^2}{f_c}&\approx& 2\left(\frac{2}{C}\right)^2\left(1+\frac{C^2x^2}{32}\right)^2\left[-dt^2+dz^2+{d\rho^2}\right]\nonumber\\ &+&\left(\frac{C}{2}\right)^2\frac{\rho^2}{\left(1+\frac{C^2x^2}{32}\right)^2} d\varphi^2 \ ,
\end{eqnarray}
where we have taken $\alpha\equiv 2f_c \rho_0^4$ as above. In Fig.\ref{fig:Ratio} we show the ratio between the function $f(x)$ computed numerically and its approximation for $\phi\gg 1$. The approximation is very good for values of $x\ge 20$.
It is worth noting that the line element  (\ref{eq:ds2far}) can be made to agree with the GR solution (up to a constant conformal factor) by just introducing a constant rescaling of $(t,z,\rho)\to (\lambda t,\lambda z,\lambda\rho)$ with $\lambda^2=64f_c/C^2$, and by suitably choosing the integration constant $\alpha$ in the corresponding definition of $h\rho^2$.

\section{Summary and perspectives \label{s-c}}

In this work we have investigated the possibility of finding analytical solutions for axially symmetric magnetic fields in the Born-Infeld theory of gravity. This type of non-asymptotically flat solutions are known as Melvin universes and are of great interest to give plausibility to the generation of pairs of entangled black holes by intense magnetic fields. In the context of GR these solutions are well known and there exist advanced solution-generating methods that allow to embed electrically charged solutions within these magnetized scenarios in a very elegant and robust way, even in the case of stationary space-times. For the Born-Infeld theory of gravity there is no guarantee that such methods can be implemented or even exist at all. For this reason, as a first step in this direction, we have investigated the very existence of Melvin-type solutions. By a suitable choice of variables (using an auxiliary line element), we have been able to write the field equations in a simple dimensionless form and obtain analytical solutions in the two regimes of interest, namely, near the symmetry axis and far from it. The complete solution can be easily worked out numerically and we have shown that our analytical approximations fit well with the exact computation. As a result, near the axis one recovers a Minkowskian geometry and far from it a standard Melvin Universe. This suggests that the embedding of electrically charged solutions, including pairs of black hole remnants and wormholes, might be possible at least in some approximate form, which motivates further research in this direction.  To investigate these questions in detail one should consider the analog of Ernst metric \cite{Ernst} in our scenario. Progress in this sense is currently underway.

\section*{Acknowledgments}

The work of C.B. and D.R.-G. was supported by NSFC (Chinese agency) grant No.~11305038 and No.11450110403, the Shanghai Municipal Education Commission grant for Innovative Programs No.~14ZZ001, the Thousand Young Talents Program, and Fudan University. G.J.O. is supported by a Ramon y Cajal contract, the Spanish grant FIS2011-29813-C02-02, the Consolider Program CPANPHY-1205388, the i-LINK0780 grant of the Spanish Research Council (CSIC), and the CNPq (Brazilian agency) grant No. 301137/2014-5.


\begin{thebibliography}{99}

\bibitem{Melvin} M. A. Melvin,
Phys. Lett. {\bf 8}, 65 (1964).
\bibitem{Wheeler}
J. A. Wheeler, Phys. Rev. \textbf{97}, 511 (1955).
\bibitem{Misner} C.~W.~Misner and J.~A.~Wheeler, Annals Phys. {\bf 2}, 525 (1957).
\bibitem{Garfinkle94}
D. Garfinkle and M. Melvin, Phys. Rev. D \textbf{50}, 3859 (1994).
\bibitem{dilaton} G. W. Gibbons and K. Maeda, Nucl. Phys. B {\bf 298}, 741 (1988).
\bibitem{axion} A. Tseytlin, Phys. Lett. B \textbf{346}, 55 (1995).
\bibitem{Higher} M.~Ortaggio,
  JHEP {\bf 0505},  048 (2005).
\bibitem{NED} G. W. Gibbons and C. A. R. Herdeiro, Class. Quant. Grav. {\bf 18}, 1677 (2001).
\bibitem{SUGRA} M. Alishahiha, B. Safarzadeh, and H. Yavartanoo,
JHEP {\bf 0601}, 153 (2006);
M. Cvetic, G. W. Gibbons, C. N. Pope, and Z. H. Saleem,  JHEP \textbf{1409}, 001 (2014).
\bibitem{Branes} R. -G. Cai and N. Ohta,
Phys. Rev. D {\bf 73},  106009 (2006).
\bibitem{Pair}  D. Garfinkle and A. Strominger,
Phys. Lett. B {\bf 256},  146 (1991);
F. Dowker, J.  P. Gauntlett, D. A. Kastor, and J. H. Traschen,
Phys. Rev. D {\bf 49}, 2909 (1994);
F. Dowker, J. P. Gauntlett, S. B. Giddings, and G. T. Horowitz, Phys. Rev. D \textbf{50}, 2662 (1994);
D. Garfinkle, S. B. Giddings, and A. Strominger,
Phys. Rev. D {\bf 49}, 958 (1994);
R. Emparan,
Phys. Rev. Lett. {\bf 75}, 3386 (1995);
S. W. Hawking and S, F. Ross,
Phys. Rev. D {\bf 52}, 5865 (1995).
\bibitem{wormholes}
M. Visser, {\it Lorentzian Wormholes: From Einstein to Hawking} (American Institute of Physics, New York, 1995)
\bibitem{Ernst}
F. J. Ernst, J. Math. Phys. \textbf{17}, 515 (1976).
\bibitem{mf1}
  P.~P.~Kronberg,
  Rept.\ Prog.\ Phys.\  {\bf 57}, 325 (1994);
  R.~Durrer and A.~Neronov,
  Astron.\ Astrophys.\ Rev.\  {\bf 21}, 62 (2013).
\bibitem{Kunze} A.~Kandus, K.~E.~Kunze, and C.~G.~Tsagas,
  Phys.\ Rept.\  {\bf 505}, 1 (2011).
  \bibitem{Hawking}  S. Hawking, Mon. Not. Roy. Astron. Soc. \textbf{152}, 75 (1971).
\bibitem{Chen} P.~Chen, Y.~C.~Ong, and D.~h.~Yeom,
  arXiv:1412.8366 [gr-qc].
\bibitem{mf2}
  D.~Grasso and H.~R.~Rubinstein,
  Phys.\ Rept.\  {\bf 348}, 163 (2001);
  A.~D.~Dolgov,
  astro-ph/0306443.

\bibitem{newref1}
  C.~M.~Yoo, T.~Harada, and N.~Tsukamoto,
  Phys.\ Rev.\ D {\bf 87}, 084045 (2013);
  R.~Takahashi and H.~Asada,
  Astrophys.\ J.\  {\bf 768}, L16 (2013).

\bibitem{Bambi}
  C.~Bambi,
  Phys.\ Rev.\ D {\bf 87}, 084039 (2013);
  Phys.\ Rev.\ D {\bf 87}, 107501 (2013);
  Z.~Li and C.~Bambi,
  Phys.\ Rev.\ D {\bf 90}, 024071 (2014).
\bibitem{or13} G. J. Olmo, D. Rubiera-Garcia, and H. Sanchis-Alepuz,
    Eur. Phys. J. C \textbf{74}, 2804 (2014).
    \bibitem{or12} G. J. Olmo and D. Rubiera-Garcia,
Phys. Rev. D \textbf{86}, 044014 (2012);
Int. J. Mod. Phys. D \textbf{21}, 1250067 (2012);
Eur. Phys. J. C \textbf{72}, 2098 (2012).
\bibitem{lor14}  F. S. N. Lobo, J. Martinez-Asencio, G. J. Olmo, and D. Rubiera-Garcia,
    Phys. Lett. B \textbf{731},  163 (2014);  Phys. Rev. D \textbf{90},  024033 (2014).
\bibitem{BIapplications} J. H. C. Scargil, M. Ba\~nados, and P. G. Ferreira, Phys. Rev. D \textbf{86}, 103533 (2012);
T. Harko, F. S. N. Lobo, M. K. Mak, and S. V. Sushkov, Mod.\ Phys.\ Lett.\ A {\bf 29}, 1450049 (2014);
M. Ba\~nados, Phys. Rev. D \textbf{77}, 123534 (2008);
M. Ba\~nados, P. G. Ferreira, and C. Skordis, Phys. Rev. D \textbf{79}, 063511 (2009);
D. N. Vollick, Phys. Rev. D {\bf 72}, 084026 (2005);
P. P. Avelino and R. Z. Ferreira, Phys. Rev. D \textbf{86}, 041501 (2012);
P. Pani, V. Cardoso, and T. Delsate, Phys. Rev. Lett. \textbf{107}, 031101 (2011);
P.~Pani and T.~P.~Sotiriou, Phys.\ Rev.\ Lett.\  {\bf 109}, 251102 (2012);
T. Delsate and J. Steinhoff, Phys. Rev. Lett. \textbf{105}, 011101 (2012);
P. Pani, T. Delsate, and V. Cardoso, Phys. Rev. D \textbf{85}, 084020 (2012).
\bibitem{BIm}
M. Born and L. Infeld, Proc. R. Soc. London A \textbf{144}, 425 (1934).
\bibitem{Deser}
S. Deser and G. W. Gibbons, Class. Quant. Grav. {\bf 15}, L35 (1998).
\bibitem{Banados} M. Ba\~nados and P. G. Ferreira, Phys. Rev. Lett. {\bf 105}, 011101 (2010).
\bibitem{Borunda} Q. Exirifard and M. M. Sheikh-Jabbari, Phys. Lett. B \textbf{661}, 158 (2008);
M. Borunda, B. Janssen, and M. Bastero-Gil, JCAP \textbf{0811}, 008 (2008).
\bibitem{Volovich} M.~O.~Katanaev and I.~V.~Volovich,
  Ann. Phys.\  {\bf 216}, 1 (1992).
\bibitem{lor-crystal} F. S. N. Lobo, G. J. Olmo, and D. Rubiera-Garcia, arXiv:1412.4499 [hep-th].
\bibitem{Torsion} G. J. Olmo and D. Rubiera-Garcia,	Phys. Rev. D \textbf{88},  084030 (2013).
\bibitem{JMMG} J. M. Martin-Garcia \url{http://www.xact.es.}

\end{thebibliography}
\end{document}